\documentclass[reprint,onecolumn,longbibliography]{revtex4-2}
\usepackage{color}
\usepackage[T1]{fontenc}
\usepackage[latin9]{inputenc}
\usepackage{geometry}
\usepackage{amsmath}
\usepackage{amsthm}
\usepackage{graphicx}
\PassOptionsToPackage{normalem}{ulem}
\usepackage{ulem}

\makeatletter

\def\tb{\textcolor{black}}

\def\my{{\mathbf{y}}}
\def\mv{{\mathbf{v}}}
\def\mu{{\mathbf{u}}}
\def\mq{{\mathbf{q}}}
\def\mvt{{\mathbf{v}_t}}

\makeatother

\usepackage{babel}
\begin{document}
\title{The waltz of tiny droplets 
and the flow they live in}
\author{S. Ravichandran}
\email{sravichandran@iitb.ac.in}
\affiliation{Nordic Institute for Theoretical Physics, KTH Royal Institute of Technology and Stockholm University, Stockholm SE 10691}
\affiliation{Interdisciplinary Programme in Climate Studies, Indian Institute of Technology Bombay, Mumbai 400076}
\author{Rama Govindarajan}
\email{rama@icts.res.in}
\affiliation{International Centre for Theoretical Sciences, Tata Institute of Fundamental Research, Bengaluru 560089, India.}
\begin{abstract}
This article describes the dynamics of small inertial particles centrifuging out of a single vortex. It shows the \tb{importance} of caustics formation in the vicinity of a single vortex: both for particle collisions and void formation. From these single-vortex studies we provide estimates of the role of caustics in high Reynolds number turbulence, and in the case of clouds, estimate how they may help in rain initiation by bridging the droplet-growth bottleneck. We briefly describe how the Basset-Boussinesq history force may be calculated by a method which does not involve huge memory costs, and provide arguments for its possible importance for droplets in turbulence. We discuss how phase change could render cloud turbulence fundamentally different from turbulence in other situations.
\end{abstract}
\maketitle

\section{Introduction}
\subsection{The flow situations}

The dynamics of finite-sized particles or droplets in turbulent flow is relevant in a wide range of natural and industrial settings. In the dispersal of pollutants due to a volcanic eruption or from industrial flue gases, in a sandstorm, or a cloud, or when a river carrying sediment disgorges itself into the ocean, we encounter such interaction. We are interested here in spherical particles, which are so small that to a very good approximation they are in Stokes flow relative to the surrounding fluid. 

Why is the dynamics, in flow, of individual Stokesian particles interesting? First, such dynamics is often not passive, in that the particle could be inertial, with finite acceleration relative to the flow. Such particles therefore may not follow fluid streamlines, and this is responsible for a variety of complex behaviour. In background turbulent flow, particle inertia leads to clustering \cite{Shaw1998}, caustics formation \cite{Falkovich2002,Wilkinson2005}, and, in some cases, coalescence and growth \cite{Wilkinson2006,Mehlig2007,Vosskuhle2014,Deepu2017,Agasthya2019} (the general features of such flows are reviewed in \cite{Balachandar2010}; see also the reviews by \cite{Shaw2003,Grabowski2013} of properties relevant to clouds). Further, the particles mentioned are often of density different from the fluid, and sediment or rise under gravity \cite{Bec2014,Ireland2016b,Ravichandran2020mammatus,Omand2020}. Notice that we use the term `particles' to stand for droplets as well, except when growth by coalescence or phase change is relevant, when we will refer to droplets. This interchangeable notation is applicable when  droplets are sufficiently small for surface tension to ensure that they remain spherical in shape.
When particles are not in dilute suspension, their interactions lead to new physics, including instabilities, see e.g. \cite{Crowley1971,Chajwa2020}. But here we discuss dilute suspensions, where inter-particle interactions can be ignored.

We are particularly interested in droplets in atmospheric clouds. Clouds are a crucial part of the Earth's energy balance. Current estimates suggest that clouds are responsible for a significant fraction of both the albedo (i.e., reflecting away of incoming solar `short-wave' radiation) of the planet, as well as  the greenhouse effect (i.e., trapping of outgoing `long-wave' radiation) \cite[see, e.g. ][]{Lohmann2005}.
The respective magnitudes of these contributions from different patches of cloud depend on their altitude, extent and thickness, as well as the droplet size distribution within. The size of each droplet in a cloud is a changing function of time, since droplets can whittle away by evaporation or grow by condensation, depending on whether their immediate neighbourhood is subsaturated or supersaturated in water vapour. The energy released during these transformations plays a crucial role in the dynamics of clouds, and in processes such as turbulent entrainment and mixing. 

The dynamics of inertial droplets are also thought to be responsible for the rapid onset of rain in warm (i.e. ice-free) clouds (\cite{Shaw1998,Kostinski2005,Falkovich2002,Wilkinson2016}). The millimetre-size droplets required for rain initiation cannot be formed either by vapour diffusion which is too slow for droplets larger than $O(10)$ micron, or by gravitational collisions-coalescence which is not sufficiently rapid for droplets smaller than $O(50)$ micron. The range $O(10-50)$ micron is called the `droplet growth bottleneck', and is thought to be `bridged' by the dynamics of inertial droplets in turbulent flow. It is known, for instance that the presence of a few \cite[one-in-a-million, see e.g. ][]{Kostinski2005} large droplets leads to a significantly more rapid generation of rain-sized droplets. We posit here that these large droplets may arise due to turbulence-induced collisions of $10$ micron droplets due to the action of small but powerful single vortices.

In the rest of this review, in \S \ref{sec:mr} we  introduce the basic equations governing the dynamics of inertial droplets and then discuss the formation of caustics due to particle inertia, neglecting higher order effects, in \S \ref{sec:onevor}. \tb{We describe the effects of a nominally higher-order correction, namely the Basset history term, in \S \ref{sec:basset}, showing that its effects are  negligible for raindrop growth due to gravitational settling}. We then discuss the effects of condensation of water vapour onto droplets on the dynamics of clouds in \S \ref{sec:cloud}, including some of our own contributions to the study of the interactions of droplet inertia, turbulence and phase change.

\subsection{The Maxey-Riley equation}
\label{sec:mr}

The dynamics of a heavy sphere in unsteady Stokes flow is described by the famous Maxey-Riley (MR) equation \cite{Maxey1983}
\begin{eqnarray}
\dot \my & = & \mv(t)
\label{posit} \\
  \dot \mv & = & -\frac{1}{St}(\mv-\mu) - \frac{3}{\sqrt{2 \beta St}}\left[\frac{1}{\sqrt{\pi t}}(\mv(0)-\mu(0)) + \frac{1}{\sqrt{\pi}} \int_0^t\frac{\dot\mv(s)-\dot\mu(s)}{\sqrt{t-s}}ds\right] + \frac{1}{Fr^2} \mathbf{e}_g
    \label{mr}
\end{eqnarray}
where $\my$ and $\mv$ respectively are the position and  velocity of the particle at time $t$, and the dots refer to differentiation in time {\em along the particle trajectory}. We prescribe that the density ratio $\beta\equiv \rho_p/\rho_f \gg 1$, where $\rho_p$ and $\rho_f$ respectively are the densities of the particle and the fluid. In this limit, the Froude number $Fr^{-2} \equiv  g T_f^2/L_f$, where $g$ is the acceleration due to gravity and $T_f$ and $L_f$ are characteristic flow time and length scales respectively. The Stokes number $St \equiv \tau/ T_f $, where 
\begin{equation}
    \tau \equiv \frac{2 \beta a^2}{9\nu},
    \label{taup}
\end{equation}
is the particle time scale, $a$ being the radius of the particle and $\nu$ the kinematic viscosity of the fluid. Equation \eqref{mr} is in general nonlinear, since $\mu=\mu(\my(t))$, the velocity of the fluid at the particle location, can depend nonlinearly on $\my$. The Faxen correction terms due to the curvature of the velocity field on the scale of the particle have been neglected in this representation of the MR equation. This is reasonable because Stokesian particles in turbulence tend to be far smaller than the structures in the flow. 
Most often \cite{Bec2003,Bec2005,Wilkinson2005} in the absence of gravity, this equation is further truncated, after the first term on the right hand side, reducing it to the simplified Maxey-Riley (SMR) equation:
\begin{equation}
   \dot\mv = -\frac{1}{St}(\mv-\mu).
    \label{smr}
\end{equation}
The Basset-Boussinesq history term \tb{(`Basset-history' or `BBH' for short)}, i.e., the second term on the right hand side of equation \eqref{mr}, has been dropped in this approximation.
At first sight, this seems to be  reasonable for particles of small Stokes number, since we have retained all terms up to $O(St)$ and neglected terms of $O(St^{3/2})$ and higher. We shall proceed with this assumption in the next section and examine it thereafter.

We ask first how turbulence, in the absence of gravity, determines particle dynamics. This question may be answered in some degree by obtaining, in detail, the dynamics near a single Lamb-Oseen vortex, as we do in Section \ref{sec:onevor} where we discuss the role and importance of single-vortex caustics. We then investigate the relevance of our findings in a generic high Reynolds number turbulence. 

\section{Dynamics near a Lamb-Oseen vortex, and the caustics question \label{sec:onevor}}

We begin this discussion by defining caustics as situations where two or more particles arrive simultaneously at the same place with different velocities. It is believed that caustics are important for enhanced particle collisions \citep[e.g.][]{Falkovich2002,Wilkinson2006,Falkovich2007} which in turn are a crucial ingredient, in the case of droplets, for growth by coalescence. Collisions are an important reason why we have rain at all, and the background turbulence in clouds,  with its associated vorticity, is considered to be a lead player in this. A natural way to divide a turbulent flow in the context of particle dynamics is into regions of vorticity and regions of strain, since inertial particles are known to centrifuge out of the former and collect in the latter \cite{Maxey1987,Squires1991}. It is well known that caustics form when particles centrifuging out of different vortices collide with each other \cite{Falkovich2002}. But are single-vortex caustics possible? This was answered in the affirmative by \cite{Ravichandran2015} and is discussed in the following. In this study we limit ourselves to particles which are far heavier than the fluid they are embedded in, but the analysis can easily be modified to include a finite density ratio, which in itself can make the dynamics more interesting. In the following, when we use the term `caustics', we are referring to single-vortex caustics.

Patches of vorticity occur in a range of shapes and sizes in turbulence, but a given patch of vorticity, if far away from other patches, will axisymmetrise into a Lamb-Oseen vortex in whose cross-section vorticity varies as a gaussian:
\begin{equation}
    \omega = \omega_v \exp\left[-\frac{r_d^2}{r_v^2}\right],
    \label{lamb-oseen}
\end{equation}
where $\omega_v = \Gamma / (\pi r_v^2)$ is the vorticity at the vortex centre, $r_d$ is the dimensional radial distance from the centre of the vortex and $r_v$ is the characteristic size of the vortex at a given time. In fact, as was pointed out by an anonymous referee of another paper, a Lamb-Oseen vortex can be termed the ``drosophila'' of turbulence, a building block from whose behaviour we may glean general principles much as one does in biology using that organism. In two-dimensional turbulence, it was verified  \cite{Ramadugu2022} that vortex patches tend to be circular and display a Gaussian cross-section of vorticity \citep[see also ][]{Moriconi2022}. There is increasing evidence in high Reynolds number simulations that intermittent vortices tend to be tubelike.

In two dimensions, in the neighbourhood of a Lamb-Oseen vortex, the SMR equation in cylindrical-polar ($r-\theta$) coordinates reads \cite{Ravichandran2015} 
\begin{eqnarray}
    \ddot r + \dot r = \frac{\zeta^2}{r^3}, \label{uni_r}\\ 
    \dot{\zeta} + \zeta = 1 - e^{-St_v r^2} \label{uni_zeta}
\end{eqnarray}
where we have nondimensionalised the equations with the special length scale
\begin{equation}
L \equiv \left(\frac{\Gamma\tau}{2\pi}\right)^{1/2},
\label{special_length}
\end{equation}
the particle time scale $\tau$, and denoted $\zeta\equiv r^2 \dot\theta $ as the non-dimensional angular momentum of the particle. Note that $L$  encompasses properties of the flow and the particle. The ratio of $L$ and the vortex size $r_v$ is also a ratio of time scales of particles and vortex, and appears in the form of a vortex time-scale based Stokes number
\begin{equation}
St_v \equiv \frac{\omega_v \tau}{2} = \frac{\Gamma \tau}{2\pi r_v^2}.
\label{eq:Stokes}
\end{equation}
Now, the Stokes number in turbulent flow is normally defined in terms of the large time scale $T_f$ in the flow, which is much bigger than the time scale of the small vortices. Particles could thus be of small Stokes number $St$ in terms of the large scale and large Stokes number $St_v$ in terms of a single small-scale vortex. We are interested in vortices in this regime.

\subsection{Point vortex, inner solution: the formation of caustics}
We note that for a point vortex, i.e., when $r_v=0$, equations \eqref{uni_r} and \eqref{uni_zeta} become parameter free. The solution in this limit describes the dynamics of Stokesian particles of any Stokes number near any point vortex irrespective of its strength. Equation \eqref{uni_zeta} is now easily solvable, and yields
\begin{equation}
    \zeta=1-(1-\zeta_0)e^{-t}.
    \label{zeta_sol}
\end{equation}
Here and in the following, a subscript $0$ indicates the value of the quantity at the initial time.
Note that $\zeta \sim O(1)$, and for the analysis below we will prescribe $\zeta_0 \ne 0$.
Equation \eqref{uni_r} supports a boundary-layer structure, and, for a point vortex, we may solve exactly for the dynamics close to, and very far away from, the vortex centre. The inner solution, for small distances from the vortex centre, is only relevant at short times, since this region is empty of particles once they get centrifuged out. Following the standard procedure for singular perturbation problems, we define inner variables 
\begin{equation}
    R_i \equiv \frac{r}{\delta_i} \quad {\rm and} \quad T_i \equiv \frac{t}{\epsilon_i}, 
    \label{inner}
\end{equation}
which are $O(1)$, with $\delta_i, \epsilon_i \ll 1$. Substituting these, and equation \eqref{zeta_sol} in equation \eqref{uni_r}, we find that the lowest order balance requires $\epsilon_i \sim \delta_i^2$. Since these are arbitrary constants defined by us, we may set $\epsilon_i = \delta_i^2$. At this order, there is no further information on what exactly $\epsilon_i$ and $\delta_i$ individually are, but this information is not necessary either. Equation \eqref{uni_r} reduces at the lowest order to
\begin{equation}
    R_i'' = \frac{\zeta_0^2}{R_i^3} \label{ri}
   \end{equation}
where primes refer to differentiation with respect to $T_i$. Equation \eqref{ri} is an autonomous nonlinear ordinary differential equation, whose solution is
\begin{equation}
    R_i^2 = \left[\frac{\zeta_0^2}{R_{i0}^2} +  {R'}_{i0}^2\right] T_i^2 + 2 R_{i0}  R'_{i0} T_i + R_{i0}^2.
    \label{caustics_time}
\end{equation}
Now caustics are formed when two sets of particles starting out at different $R_0$ arrive at the same radial location $R_c$ at a particular time $T_c$, with the subscript $c$ denoting caustics. The caustics time for a given set of initial conditions may be obtained from equation \eqref{caustics_time}. For example, if two particles, initially at $R_{01}$ and $R_{02}$ have the same $\zeta_0$ and start out with zero radial velocity, we have
\begin{equation}
    T_c=\frac{R_{01}R_{02}}{\zeta_0 }.
    \label{tc}
\end{equation}
We arrive at the conclusion that the inner solution supports the formation of caustics for a point vortex. Notably, all terms in equation \eqref{tc} are $O(1)$, which means that the caustics time $t_c=\epsilon_i T_c$ is very short. 
\subsection{Outer solution: no caustics}
\begin{figure}
    \centering
    \includegraphics[width=0.46\columnwidth]{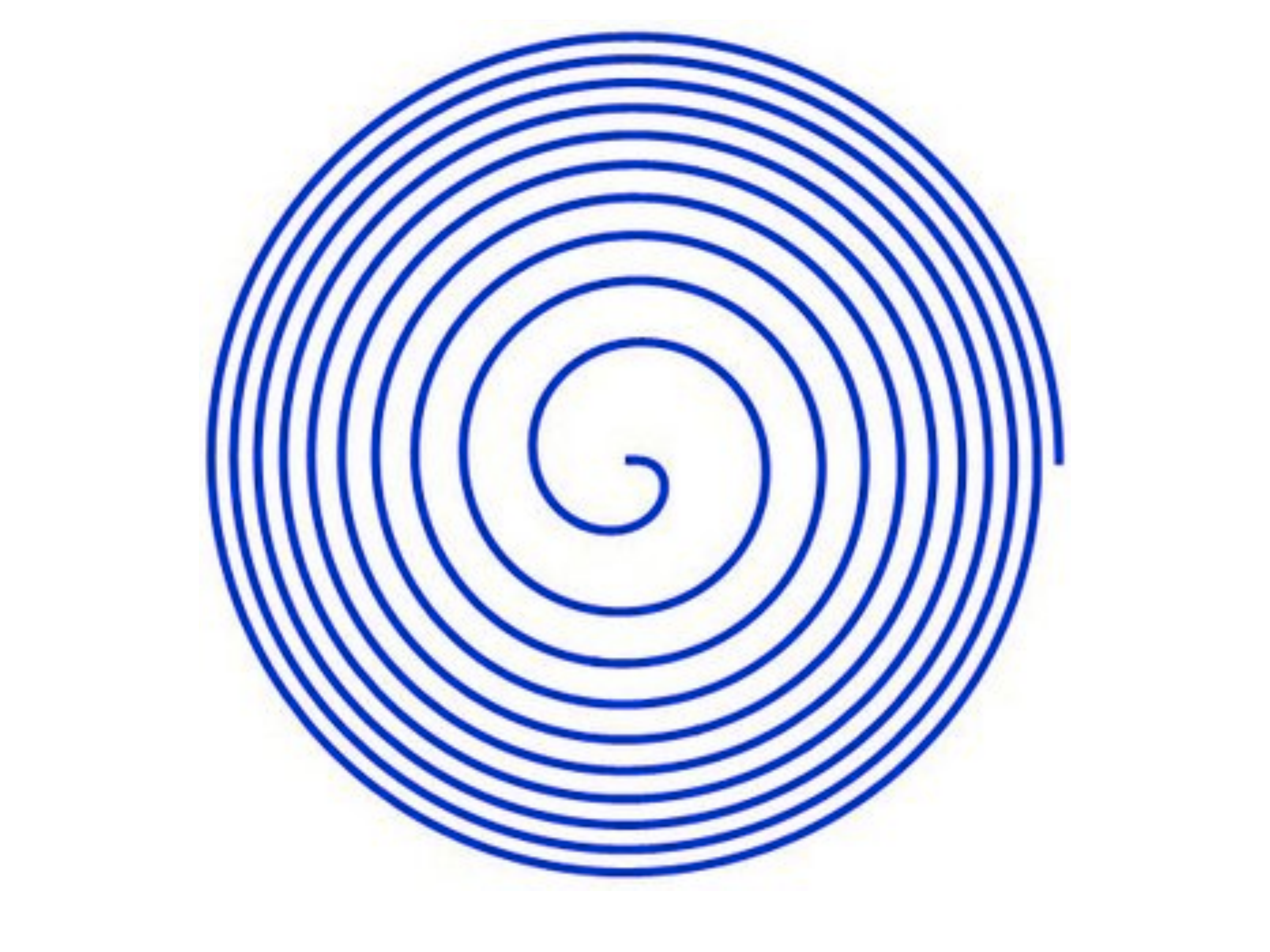}
    \caption{Fermat's spiral describing the centrifugation of a heavy particle in the outer region of a vortex. As time goes on, the radial distance covered per cycle reduces and the time taken to cover each cycle increases.
    }
      \label{fig:fermat}
\end{figure}
At large time, all particles evacuate the vicinity of the vortex, and occupy the region at large $r$. Moreover equation \eqref{zeta_sol} gives $\zeta=1$. The outer equations may be obtained in a similar manner as above by defining outer variables, this time as $R_o=\delta_o r$ and $T_o = \epsilon_o t$ where $\delta_o,\epsilon_o \ll 1$ to get, at the lowest order,
\begin{equation}
 R_o' = \frac{1}{R_o^3}, 
    \label{outer}
\end{equation}
the primes now referring to differentiation with respect to $T_o$. Note that the subscript $o$ stands for the outer equation, and is not to be confused with the subscript $0$ for initial conditions. The balance this time yields $\epsilon_o=\delta_o^4$. Equation \eqref{outer} may be solved immediately to yield $R_o^4 - R_{o0}^4 = 4T_o$. At large times $R_o \gg R_{o0}$, so we may neglect the latter in comparison to the former. When put back in the physical variables, we have  $r^4=4t$, with $r^2\dot\theta=1$. In other words $\theta/r^2=1/2$, i.e., a particle far away from the vortex will execute a Fermat's spiral, shown in schematic in figure \ref{fig:fermat}, thus centrifuging ever more slowly out of the vicinity.

It is evident that from equation \eqref{outer} that a given radial distance away from the vortex is associated with a unique particle radial velocity $\dot R_o$. This means that particle dynamics may be described in the outer region in terms of a field for the particle velocity, namely $\dot R_o=\dot R_o(R_o,\theta,T_o)$. This in turn means there is no possibility of caustics in this region. 

\subsection{The complete solution for single-vortex caustics}
\label{sec:caustics}

The singular perturbation study above is carried out just to be instructive. The complete SMR for a particle near a Lamb-Oseen vortex, equations \eqref{uni_r} and \eqref{uni_zeta}, are trivial to solve numerically and we have done so. From the lowest-order singular perturbation theory, we saw for a point vortex that caustics happen at early times and $r \ll 1$ and never at late times and $r \gg 1$. We may thus expect that the limit for caustics will happen at a caustics radius $r_c \sim O(1)$. In fact, for a range of initial conditions \citep{Ravichandran2015} we find that
the caustics radius is given by
\begin{equation}
    r_{cd}^2 \sim \Gamma \tau, \quad {\rm or} \quad r_c \sim 1,
    \label{rc}
\end{equation}
where the subscript $d$ denotes a dimensional quantity. 

\begin{figure}
    \centering
    \includegraphics[width=\columnwidth]{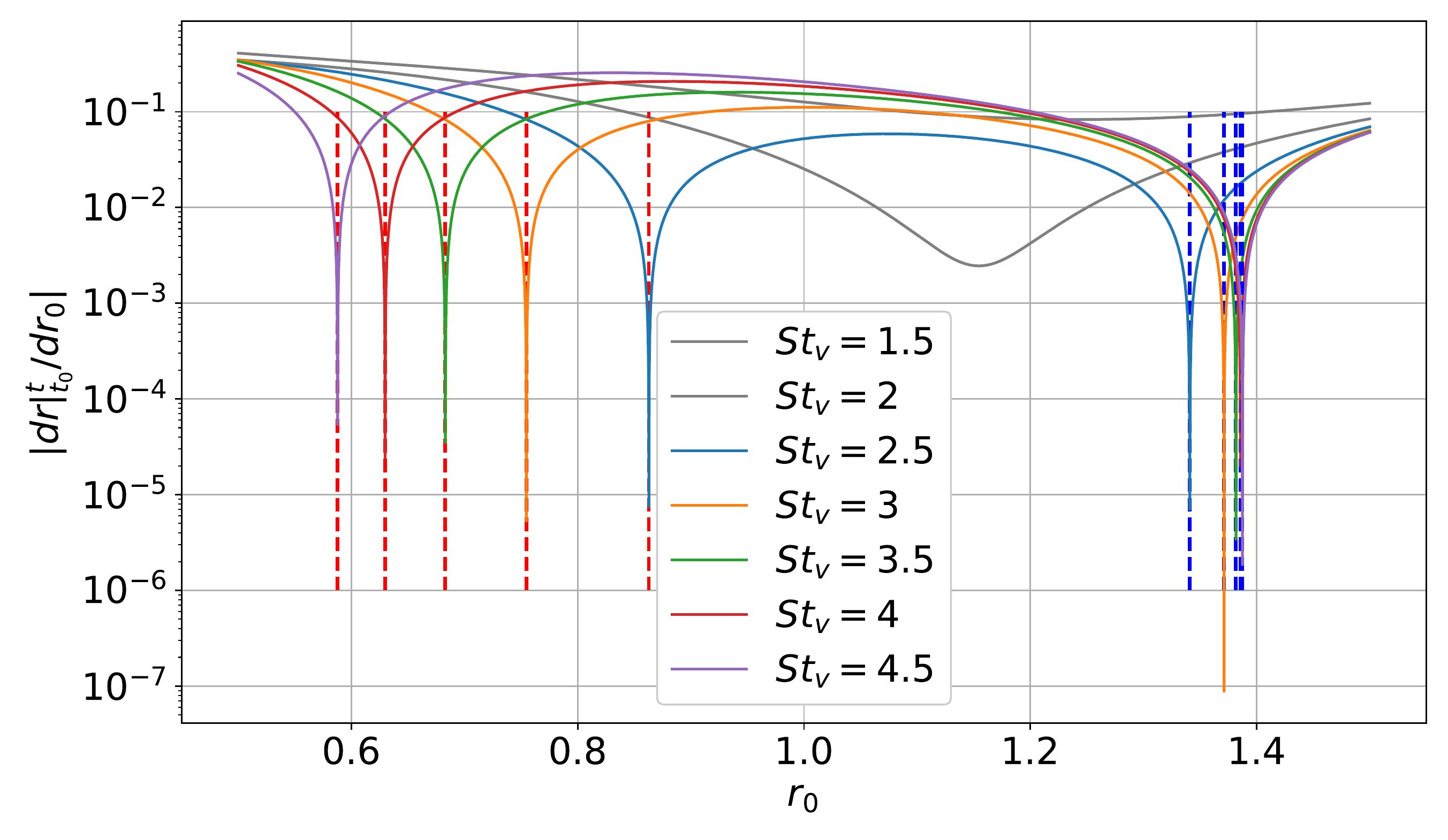}
    \caption{\label{fig:St_vs_caustics_region} The range of initial locations (the region between the red dashed lines, where $r_0=r_{ci}$ and the blue dashed lines, where $r_0=r_c$)  broadens as the Stokes number $St_v$ (equation \ref{eq:Stokes}) increases. For each curve, caustics occur in locations bounded by these vertical asymptotes. Thus, for $St_v\lesssim 2.3$, no caustics occur. These curves were obtained using inertial finite-time Lyapunov exponents (iFTLEs, see \cite{Sudharsan2016} in the radial direction, the values of which are plotted as the ordinate.}
\end{figure}
\begin{figure}
    \centering
     \includegraphics[width=0.46\columnwidth]{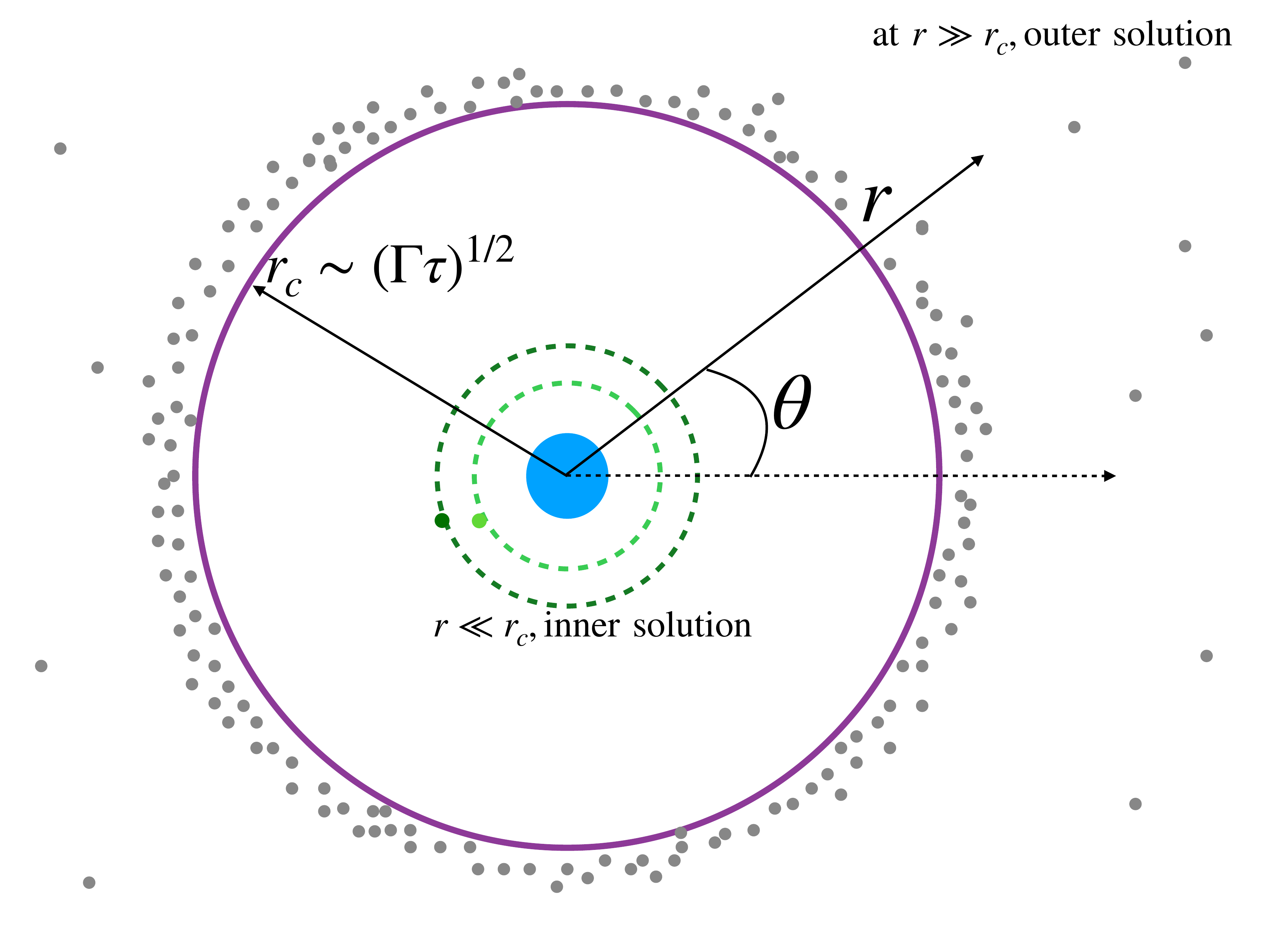}
    \label{fig:caustics}
    \caption{Schematic of single-vortex caustics. The vortex is shown by the filled blue circle. Caustics can happen only inside $r_c$, and an example of an inner light green ring of particles just about to overtake the outer dark green ring is shown. Particles within the caustics radius $r_c$ get centrifuged out very rapidly, thus evacuating this region. Just beyond, an intensely clustered ring of particles collects and continues centrifuging out at a far slower rate. This is a singular perturbation problem with distinguished limits at radii well within and well outside the caustics radius.  }
\end{figure}
For a Lamb-Oseen vortex of finite size too, the complete solution of the SMR equation shows caustics, see figure \ref{fig:St_vs_caustics_region}. The difference is that while for a point vortex, particles starting from any $r<r_c$ participated in caustics, for the Lamb-Oseen vortex we have caustics occurring for particles starting within $r_{ci} \le r \le r_c$, where $r_{ci}$ is an inner limit for caustics formation, which decreases as the Stokes number $St_v$ increases, as seen from figure \ref{fig:St_vs_caustics_region}. Further, all caustics happen at times of $O(\tau)$, and the region within $r_c$ is devoid of particles at later times, irrespective of the value of $r_{ci}$. In other words, the region $r_d \lesssim (\Gamma \tau)^{1/2}$ where caustics form is also the region which gets voided extremely quickly (on the particle timescale) of particles due to rapid centrifugation. This will be important for our calculations below of void volumes. Another finding is that in the case of single-vortex caustics, there is a direct and strong correlation between collisions and the formation of caustics. Collisions are overwhelmingly the result of caustics formation, and so caustics are crucial for droplet growth. Figure \ref{fig:St_vs_caustics_region} indicates in schematic that there is a sharp increase in the number density of particles just outside $r_c$, which is one reason for enhanced collisions. This highly clustered band then centrifuges out slowly on the Fermat spiral time scale. Incidentally we do have rare occurrences of collisions without caustics, when the centres of two particles come closer together than the sum of their radii. More details are available in \cite{Ravichandran2015,Deepu2017}. 

\begin{figure}
    \centering
    \includegraphics[width=1\columnwidth]{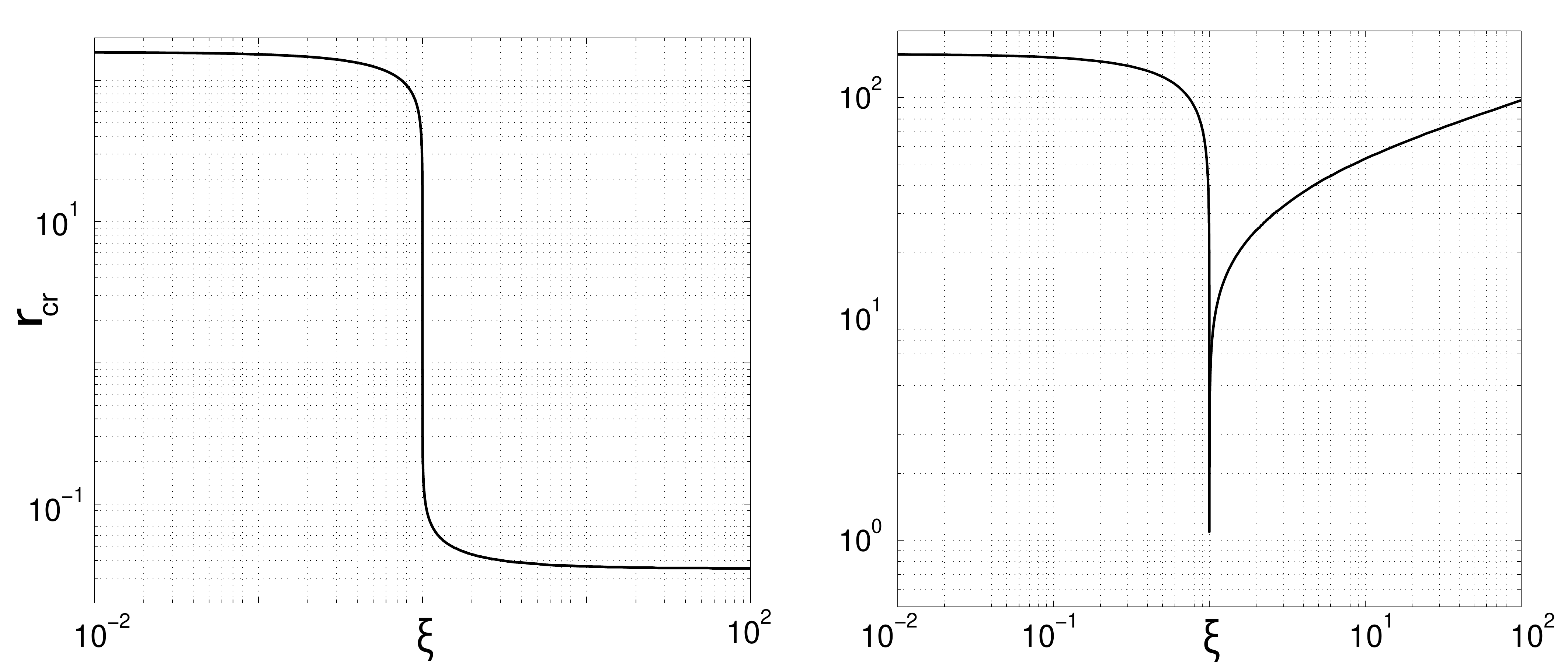}
    \caption{\label{fig:caustics_bidisp} Nondimensional polydisperse caustics radii as a function of the ratio of time scales $\xi = \tau_2 / \tau_1$  of outer to inner particle. Small differences in size lead to large variations. Left: particles initially following circular streamlines; right: particles initially stationary. See text for a discussion of relevance.}
\end{figure}

The calculation of the caustics time in equation \eqref{tc} was made for two heavy spherical particles of {\em identical} size and therefore equal time constant $\tau$. In the case of two particles with different $\tau$, equation \eqref{caustics_time} can be rewritten in dimensional form for each particle, and the caustics radius and time calculated. In Ref. \citep[][ figure 3 therein]{Deepu2017}, we showed that even very small differences in the Stokes numbers can vastly change the ``polydisperse caustics radius'', denoted here by $r_{cp}$ and quantified in figure \ref{fig:caustics_bidisp}. The particle starting at a smaller distance from the vortex centre is denoted by the subscript $1$ and the other one by $2$. If the outer particle is larger, the initial conditions determine whether the smaller one can overtake it. The differences in figures \ref{fig:caustics_bidisp} between the two initial conditions therefore merits further study. When the inner particle is larger, $r_{cp}$ is larger by two orders of magnitude than $r_c$. At first sight, it appears that our findings in the earlier section are thus rendered irrelevant. This is emphatically not the case: our conclusions hold for a distribution of particle sizes, unless this distribution is extremely wide. The relevant caustics radii are still the individual $r_c$'s. Well beyond this, each particle follows its outer solution of executing an ever-more slowly widening Fermat spiral, and approach velocities between particles, and therefore collision rates, are extremely low. Nevertheless, $r_{cp}$ is technically the caustics radius for polydisperse particles. A tangible effect of polydispersity is that the bigger droplet's radial velocity being larger than the smaller one's, it can collide with and collect more droplets rather easily, and so polydisperse caustics afford a ``rich get richer'' scenario, by which droplets which are already larger are strongly favoured to grow more easily.

We shall use these findings on dynamics near one vortex in the next subsection to estimate particle collision due to caustics in high Reynolds number turbulence. We shall see that vortices of smaller and smaller scale contribute more and more significantly to the total number of caustics events.

\subsection{An estimate of single-vortex caustics in high Reynolds number turbulence}
A vast amount of new knowledge is being accrued about high Reynolds number turbulence under homogeneous isotropic conditions, though these do not include the thermodynamics of phase change. As we shall see in section \ref{sec:cloud}, due to the heat released during condensation, small-scale vorticity is likely to be more prevalent in a cloud than in the common homogeneous isotropic turbulence, but our analysis here based on `normal' turbulence will provide a possibly conservative estimate of droplet growth due to caustics near a single vortex. For this we rely heavily on the excellent study by Buaria and Pumir \cite{Buaria2022}, showing how extreme vorticity follows universal behaviour at high Reynolds number. They have studied Taylor microscale-based Reynolds numbers $R_\lambda$ up to $1300$, but their demonstration of universality allows for extrapolation to higher $R_\lambda$. Such extrapolation is of course subject to validation against future simulations. The vorticity field is known to be far more intermittent than the strain field, and we shall find this to be very important. Following \cite{Buaria2022}, we define $\Omega \equiv \omega_i \omega_i$. At the small scales, we have \cite{Buaria2022}
\begin{equation}
    \frac{r_v}{\eta_K} \simeq (\Omega t_K^2)^{-\gamma/4} \quad \simeq \left(\frac{t_K \Gamma}{r_v^2}\right)^{-\gamma/2} 
    \label{eta}
\end{equation}
where $\eta_K$ and $t_K$ are the Kolmogorov length and time scales, $u_v$ is a typical velocity of a vortex of size $r_v$, and $\gamma$ is a parameter obtained from direct numerical simulations. Here $\Gamma \sim r_v u_v \sim \Omega^{1/2} r_v^2$, where $u_v$ is a velocity scale of the vortex. As $R_\lambda \to \infty$, it is expected that $\gamma \to 1$, but in our range of interest, we may take $\gamma \simeq 0.8$.

What follows is an order of magnitude calculation for vorticity at small scales. For caustics to form, our single-vortex calculations have shown that we must have the caustics radius $r_c$ to be larger than the vortex radius $r_v$. Using equations \eqref{eta} and \eqref{rc}, we have the simple requirement for caustics, that
\begin{equation}
    \Omega \geq \frac{1}{\tau^2}, \quad {\rm or} \  \Omega_{min} = \frac{1}{\tau^2},
    \label{Omega_min}
\end{equation}
where $\Omega_{min}$ is the squared vorticity amplitude of the largest sized vortex that can give rise to caustics. In other words, vortices of $St_v \ge 1$, i.e., of time scale equal to or faster than the particle time scale participate in caustics. This estimate is exactly in line with the observations of \cite{Motoori2022}, in direct numerical simulation of particulate channel flow, that significant voids due to centrifugation are formed for vortices whose turnover time is comparable to the particle timescale. From equation \ref{eq:Stokes}, this is equivalent to
$$
\frac{2\pi}{\omega_v} = \tau_p, \text{or } St_v = \pi,
$$
which is comparable to the critical Stokes number of $2.3$ for caustics to occur (see figure \ref{fig:St_vs_caustics_region}). Our contention is that what \cite{Motoori2022} see in their simulations is a consequence of the caustics radius, and of \eqref{Omega_min}.

As the  Reynolds number increases, so does the intermittency in $\Omega$ \citep{Buaria2022}, and the fastest, or most extreme, time scale $t_{ext}$ in the flow reduces, according to 
\begin{equation}
    t_{ext} \sim t_K R_\lambda^{-\xi},
    \label{tmin}
\end{equation} 
where $\xi$ is an $O(1)$ number, changing very slowly with $R_\lambda$. Again, for our purposes, $\xi \simeq 0.8$.
Buaria \& Pumir show that when $\Omega$ is scaled by $t_{ext}^{-2}$, the tails of its probability density over a range of $R_\lambda$ from $140$ to $1300$ collapse into a single curve.   

To simplify our estimate, we assume all particles are of uniform size. The void area $a_{void}$ around a vortex, within which there are no particles, is comparable to the area of the caustics region, i.e.,
\begin{equation}
 a_{void} \sim  \Gamma \tau  \sim u_v r_v \tau \sim \Omega^{1/2} r_v^2 \tau \sim x^{(1-\gamma)/2} \eta_K^2 \frac{\tau}{t_K},
 \label{avoid}
\end{equation}
where we have used equation \eqref{eta}, and defined $x \equiv \Omega t_K^2$. Knowledge of the probability density $p(x)$ of $x$ enables us to obtain the void fraction due to particle evacuation in the flow. Due to homogeneity, we can obtain the void area fraction on any plane, and this would be the same fraction in other planes, and therefore in the entire volume. Assuming stationarity, $p(x)\Delta x|_{x1}$ is the fraction of the volume of the flow which is occupied by vortices with $x$ in the range $x_1$ to $x_1+\Delta x$. Similarly on any planar cross section in the flow the fraction of area $A(x)$ occupied by vorticity in the corresponding range would be $p(x) \tb{\Delta x_{x1}}$. This area is split into an average of $A(x)/r_v^2$ vortices, where $r_v^2$ is the area of a vortex in order of magnitude. On a plane of unit area, since
\begin{equation}
    \int_0^\infty p(x) dx =1, \quad {\rm we \ have} \quad
    \int_0^\infty A(x) dx =1,
\end{equation}
 and the void fraction may be obtained, using equation \eqref{avoid}, as 
\begin{equation}
    f_{void}= \int_{x_{min}}^\infty \frac{p(x)}{r_v^2} a_{void}(x) \ d x \quad = \frac{\tau}{t_K} \int_{x_{min}}^\infty p(x) x^{1/2}  \ d x.
    \label{fvoid}
\end{equation}
Note the lower limit of the integral, $x_{min}=\Omega_{min}t_K^2$, the minimum $x$ for caustics, below which no void is created.
This void fraction is attained on very fast time scales, and may be taken as the instantaneous fraction for a given vorticity field.

The idea is to find the squared amplitude of vorticity $\Omega_{min}$ beyond which vortices will participate in caustics, and then use equations \eqref{avoid} and \eqref{fvoid} to obtain the fraction of volume in the turbulent flow occupied by voids. A small fraction of the droplets which have evacuated these voids would participate in caustics events, giving opportunity to coalesce and make bigger drops. It is an oft quoted estimate of \cite{Kostinski2005} that all we need for a runaway growth of droplets into raindrops is a very small minority of larger drops, of the order of one in a million drops. This is consistent with our finding of ``rich getting richer'' by single vortex caustics. Using this, we may estimate the generation of caustics by intermittent vortices in turbulent flows. This exercise is carried out in the following section to give us an estimate of the minimum turbulence levels which can produce rain.

\subsection{Cloud estimates}
\label{sec:estimates}

\begin{figure}
    \centering
    \includegraphics[width=0.50\columnwidth]{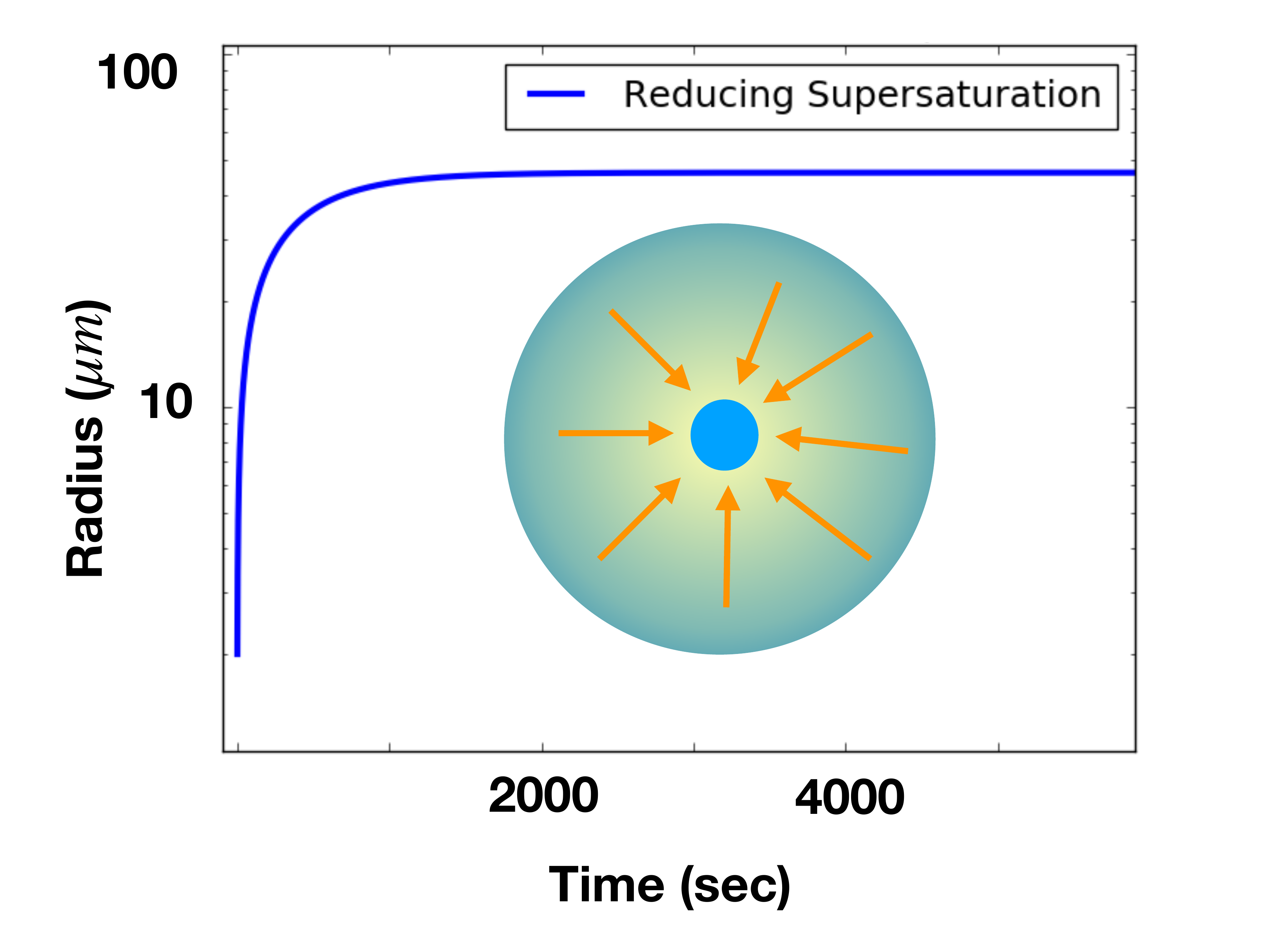}
     \includegraphics[width=0.48\columnwidth]{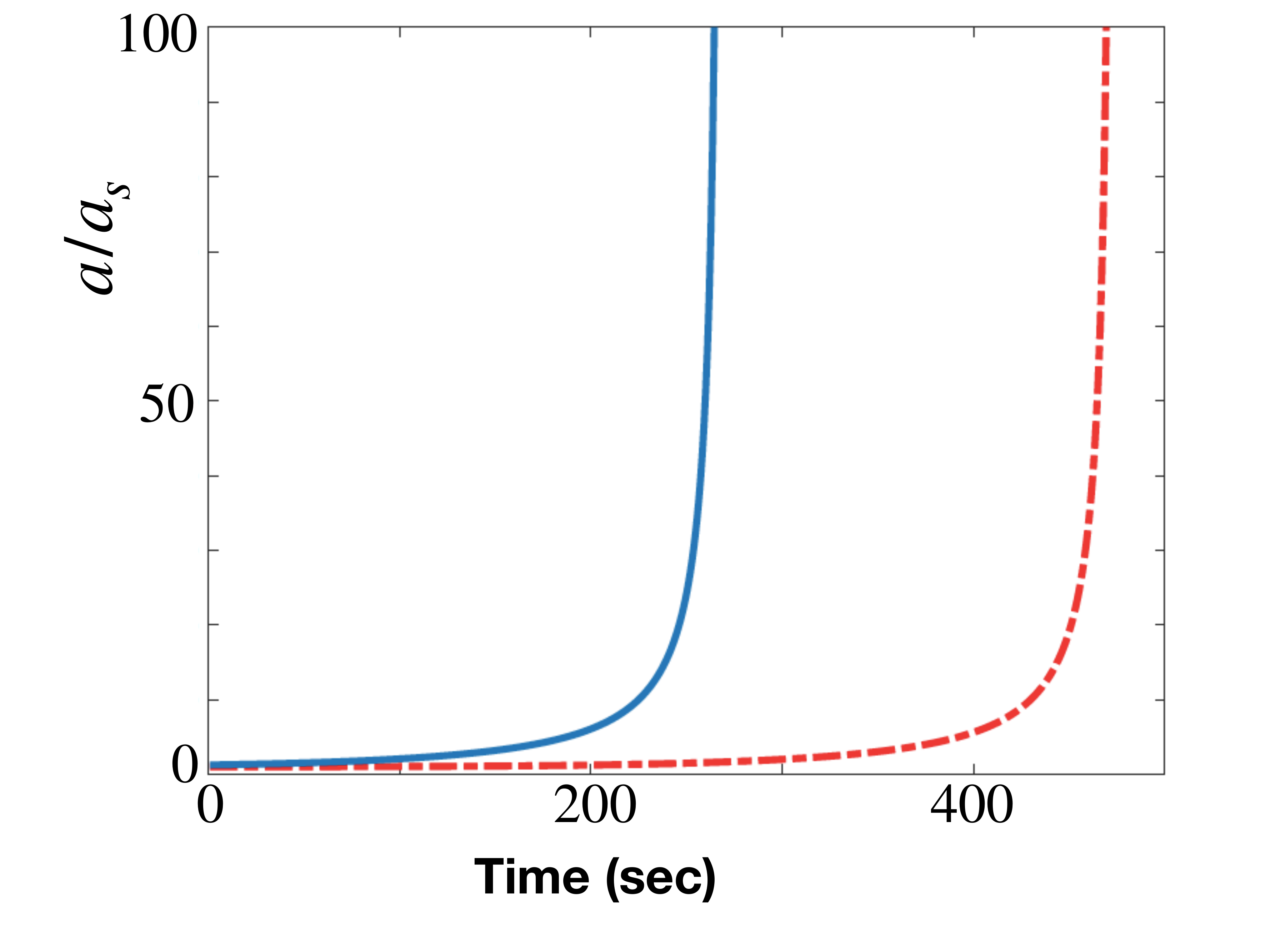}
    \label{growth}
    \caption{Droplet growth (a) by condensation in a quiescent environment supersaturated by $5\%$ in water vapour, and (b) by gravitational settling and the associated collisions and coalescence. The initial radius of the larger drop is $2^{1/3}a_s$ for the blue solid line and $1.01 a_s$ for the red dashed line. With $a_s=10$ microns, $N=10^9/m^3$ and $F=1$, and air-water properties, the time shown is in seconds.
    }
    \label{condensation_gravity}
\end{figure}

In cloud physics we often encounter the term `droplet growth bottleneck'. This is outlined as follows. A nascent cloud consists of large numbers of cloud condensation nuclei (CCN), typically aerosol particles of the order of 1 micron in size, in a background slightly supersaturated in water vapour. Water vapour at these concentrations can only condense into liquid water on CCNs, and will not undergo spontaneous condensation in their absence. It is known that diffusion of water vapour and condensation onto CCNs results, in a very short time, in cloud droplets of size $\sim 10$ microns. But beyond this size, their growth by condensation is too slow to create raindrops, as seen in figure \ref{condensation_gravity}(a). This estimate is made accounting for radial diffusion of water vapour onto to a droplet in a background which is 5 percent supersaturated, allowing for the reduction in supersaturation near the droplet caused both by condensation and the resulting increase in temperature. We mention that this level of supersaturation in higher than seen in most clouds. A falling droplet of instantaneous radius $a$, which is larger than the average radius $a_0$, has another growth mechanism accessible to it: being larger, it will sediment faster than the smaller droplets, since its dimensional terminal velocity 
\begin{equation}
   \mv_{td}= g \tau_p {\mathbf e}_g, \quad {\rm or \ in \ nondimensional \ terms,}  \quad \mv_t=-\frac{St}{Fr^2}.
   \label{terminal}
\end{equation}
From this, and equation \eqref{taup}, we see that the bigger drop sediments faster than the average by a factor $(a/a_0)^2$, and can therefore participate in what we would call ``gravity induced caustics'' events, i.e., catch up with and overtake smaller droplets. The justification for assuming that a droplet always falls at its terminal velocity is given as follows. If we drop the history term in \eqref{mr} but retain the gravity term, this equation may be readily solved, as is often done in undergraduate fluid mechanics, to give
\begin{equation}
  \mv  = \mvt \left[1-e^{-t/St}\right] + \mv_0 e^{-t/St}.
    \label{exponential_vt}
\end{equation}
In a short time, of $O(St)$, the particle attains its terminal velocity, and we may assume this process to be instantaneous.

Consider a large three-dimensional box of quiescent fluid, with $N$ identical small droplets per unit volume of size $a_s$ uniformly distributed in space. The subscript $s$ stands for `small'. The suspension is dilute, so the droplets do not interact with each other. Equation \eqref{exponential_vt} tells us that all these droplets are sedimenting steadily at their terminal velocity ${\mathbf v}_{ts}$. Now consider one larger droplet of radius $a$, slightly bigger than $a_s$. This droplet falls faster than the smaller ones, and will thus collide with the smaller droplets below it, which are within a cylinder of radius $a+a_s$ from the centre of the large drop. We take it that a fraction $F$ of collisions result in coalescence, allowing the bigger drop to keep growing. Since the suspension is dilute, we may take it that the mean collision time is far greater than the time scale of the big droplet. Using equation \eqref{terminal}, the growth of the larger droplet is then given by (from the collision rate \cite[][in, e.g., equation 2 thereof]{Wilkinson2016})
\begin{equation}
    \frac{dy}{dt} = \frac{\pi F N g \tau_s a_s^2} {3} \frac{(y-1)(y+1)^3}{y^2},
    \label{agrow}
\end{equation}
where $y=a/a_s$.
Figure \ref{condensation_gravity}(b) shows that the growth due to gravity by the above equation, assuming every collision results in coalescence, i.e. $F=1$. The smaller droplets are taken to be $10$ microns in radius, with a high number density of $N=10^9/m^3$, which corresponds to cumulonimbus clouds. The droplet size is seen to diverge at a finite time, but we note that that when the drop is bigger than $\approx 50$ micron \cite[e.g.][]{Good2014}, it will no longer be in Stokes flow, and equation \eqref{agrow} would cease to be valid. If two smaller drops coalesce to form the initial large drop, its radius would be $2^{1/3}a_s$. Such a drop would grow to $10a_s$ in about 4 minutes, whereas a drop initially at $1.01a_s$ would take twice that long. With a smaller $N$, and with $F<1$, the growth rate would be proportionately slower. Since rain is estimated to happen within about ten minutes of the initiation of condensation, and unlike assumed above, collision efficiencies for these small droplets are $F \ll 1$ \cite[][quoting Pruppacher and Klett]{Wilkinson2016},  such growth is too slow to create raindrops. We caution the reader again that the analysis above is constrained by the applicability of the SMR equations \ref{mr}. Large droplets have larger drag forces acting on them with the drag increasing as $|\mv|^2$ for sufficiently large droplets. By accounting for finite Reynolds number effects, it has been shown by others that droplets which are already of size 50-100 microns can grow rapidly by gravitational collisions and coalescence to millimetre size. 

\begin{figure}
    \centering
    \vskip-0.2in
     \includegraphics[width=0.90\columnwidth]{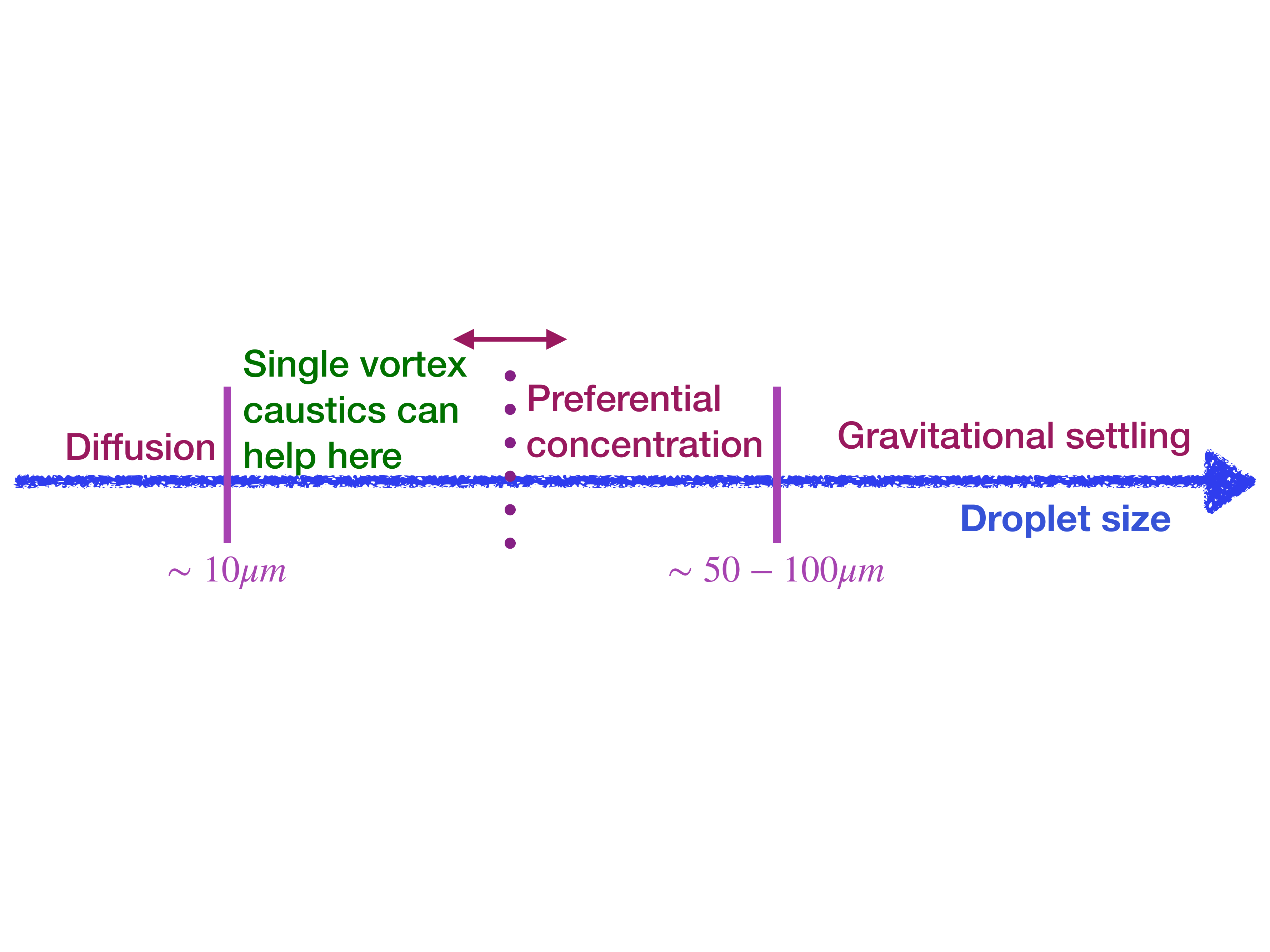}
      \vspace{-1.3in}
    \caption{\label{fig:bottleneck_schematic} Schematic of the droplet-growth bottleneck in clouds. Droplets smaller than about 10 microns grow by diffusion and condensation, and those bigger than 50-100 microns by gravitational settling and the associated collisions and coalescence. Turbulence is said to bridge this gap. While growth in the higher end of this size range could be enhanced by preferential concentration, we propose single-vortex caustics as being effective near intermittent vortices for the smaller end of this range.
    }
\end{figure}
We are thus confronted with the `droplet growth bottleneck' question, which asks how droplets grow rapidly from about 10 to 50 microns. Preferential concentration in strain-dominated regions in turbulence is often presented as the answer. But in simulations of turbulence, such preferential concentration has only been seen to take place for particles of Stokes number of order unity, based on  the Kolmogorov timescale. Cloud droplets of $10$ micron radius typically have Stokes numbers much smaller than unity in this scale. Figure \ref{fig:bottleneck_schematic}  summarises this problem. Below, we make a case for single-vortex caustics to be effective in the early part of the growth process beyond the diffusion range.

We make estimates for droplet collisions due to caustics in a cumulonimbus cloud, assuming that cloud turbulence is unchanged by phase change, and assuming that the collapse of the tails of the vorticity probability density seen by \cite{Buaria2022} may be extended to higher $R_\lambda$. 
In a cumulonimbus cloud we have $Re \simeq 10^8$, i.e., $R_\lambda \simeq 10^4$, and turbulent dissipation $\epsilon \simeq 0.1 m^2/s^3$ \cite{Grabowski2013}.  This yields the large length and velocity scales to be $L_f \simeq 300m$ and $U_f \simeq 3m/s$, and Kolmogorov scales $\eta_K \simeq 3\times10^{-4} m$, $u_K \simeq 0.03 m/s$, and therefore $t_K \simeq 0.01 s$. We shall see below that this mechanism is likely to be active in clouds of lower turbulence levels as well.

\begin{table}
\centering
\begin{tabular}{|c|c|c|c|c|}
\hline
$R_\lambda$     & $t_K$ & $\eta_K$ & $x_{min}=\Omega_{min}t_K^2$ & $f_{void}$ \\
\hline
$650$ & $0.1 s$ &  $10^{-3}$ m &  $2500$ & $ \sim 10^{-10}$ \\
$1300$ & $0.027 s$ & $5 \times 10^{-4}m$  & $ 200$  & $ \sim 10^{-4}$ \\
$10^4$ & $10^{-2}s$ & $3 \times 10^{-4}m$ &  $25$ & $O(1)$  \\
\hline
\end{tabular}
\caption{\label{tab:fvoid} Estimates of squared vorticity amplitude of the vortex of the largest time scale that can participate in caustics, and the volume fraction devoid of droplets as functions of the Taylor microscale Reynolds number. Turbulence data provided in Fig.2a of \cite{Buaria2022} is used, along with extrapolation to large $R_\lambda$ as suggested in that study. Calculations are based on a particle time scale of 2 milliseconds, corresponding to a water droplet of $10^{-6} m$ radius in air.}
\end{table}
The probability density function for squared vorticity amplitude is provided in Figure 2a of \cite{Buaria2022}. We use this to generate the estimates shown in Table 1, for three values of $R_\lambda$. Consistent with estimates for turbulent dissipation $\epsilon$ provided for clouds by \cite{Grabowski2013}, we prescribe $\epsilon=0.01$ and $0.001m^2/s^3$ for $R_\lambda=1300$ and $650$ respectively. With these, and assuming constant dissipation for large scales, (down to the Kolmogorov scale), we obtain the Kolmogorov time and length scales shown in the table. Now a $10$ micron droplet in a cloud has a time-scale $\tau \simeq 2$ milliseconds. Using equation \eqref{Omega_min}, we obtain the minimum squared vorticity amplitude, above which caustics can form, and display these in the table. We note that all vortices which can form single-vortex caustics in a cloud are far smaller than the Kolmogorov scale, i.e., they are intense intermittent vortices. This points to the need for simulations which resolve well below the Kolmogorov scale, up to $\eta_{ext}$. The highest $R_\lambda$ for which such well-resolved direct numerical simulations have been performed in homogeneous isotropic turbulence is $1300$. Already at this Reynolds number a large fraction of all vorticity is contained in the intermittent range $\eta_K \ge \eta > \eta_{ext}$. From fig. 2a of  \cite{Buaria2022}, and using equations \eqref{avoid} and \eqref{fvoid}, we can obtain the fraction of the flow devoid of particles for $R_\lambda=650$ and $1300$. The case of $R_\lambda=10^4$, consistent with deep convective clouds, is treated differently. Since direct numerical simulations are not available, we can get at best a crude estimate. We assume here that the collapse of the probability density curves seen in figure 3a of \cite{Buaria2022} extends up to this Reynolds number. Under this assumption, we have $p(x \times t_K^2/t_{ext}^2)|_{10^4} = p(x^* \times t_K^2/t_{ext}^2)|_{1300}$, yielding $x^*/x=0.005$. This provides us the transformation to use the probability density function of $R_\lambda=1300$ to obtain an estimate for $R_\lambda=10^4$. But there is a source of error here. The collapse of data is valid only for the tails of the distribution, whereas $x^*$ in this case works out to be $0.131$. Secondly, the data for low $x^*$ are not easy to read off in the data of \cite{Buaria2022}. To make allowances for this we obtain only an order of magnitude estimate for void fraction at $R_\lambda=10^4$, as being of $O(1)$. Our scaling leads us to expect that a significant fraction of all droplets at this Reynolds number will be participating in caustics events, leading to rapid rainfall initiation. But indeed nothing short of actual cloud simulations which are resolved well past the Kolmogorov scale at this high Reynolds number will provide us the real estimate. In particular, the collapse with $R_\lambda$ seen up to $1300$ may not persist up to another order of magnitude increase.

The presence of voids in the spatial distribution of inertial particles is well-known \cite{Bec2007a,Bec2007b}. In particular, Bec \& Ch\'etrite \cite{Bec2007b} suggest a model in which vortices eject particles of timescales comparable to their turnover times, and  show that this model is able to reproduce observed statistics of spatial distributions of particles. Our ideas are similar, except that our calculations are based on an exact model for the flow and its coupling with particle inertia.

Clouds are not always this turbulent and an  $R_\lambda$ of $1300$ may not be unrealistic in some cloud situations. So we restrict our further discussion to lower Reynolds numbers. In a volume of a cubic meter within a cumulonimbus cloud, there are $O(10^9)$ droplets of 10 micron radius, so by our estimate, $10^5$ drops per cubic metre may be termed as caustics droplets. In other clouds this may be an order of magnitude lower, since $N$ is lower. We have found \citep{Ravichandran2015,Deepu2017} that caustics droplets 
that do undergo collisions do so on time scales comparable to the inertial particle timescale $\tau$, and within a distance comparable to the caustics radius $\sqrt{\Gamma \tau}$ of the vortex.
Allowing for a collision efficiency $F$ of $0.01-0.1$ we expect about 1 in every hundred thousand or a million droplets to coalesce with another, and become a factor of $2^{1/3}$ larger in radius than average. These larger drops have a correspondingly higher Stokes number, and larger caustics radius, so they can be thrown out further and attain higher momentum than the average droplet, with  enhanced probability for further collisions and growth due to both turbulence and gravity. The argument, following Kostinski \& Shaw  \cite{Kostinski2005}, is that it is the initial collisions that are crucial in determining the `luck' of the droplets that ultimately become rain, since the later collisions become progressively more likely. Therefore, a mechanism such as ours that explain the initial collisions can be a crucial part of the story.
We may conclude that, by this mechanism, single-vortex caustics are extremely unlikely to initiate rainfall at $R_\lambda$ of $650$, but at $1300$ and beyond, they can initiate rain, with increasing rapidity for higher Reynolds numbers.

\section{Basset-Boussinesq history \label{sec:basset}}

In the above we had worked with the SMR equation, whereas particle dynamics are described by the MR equation \eqref{mr}. We notice that the next term in the power series of $St$ is the \tb{Basset-history} force, being a half-power of Stokes smaller than the Stokes drag term. This term is usually neglected, not because we are convinced about its smallness of magnitude, but simply because it is too hard to compute, if one were to follow the naive approach of performing the integral at every instant of time for every particle \cite{Haller2019}. It is clear that this approach is not feasible for any large flow. For example a cloud contains a billion droplets per cubic metre, and performing this integral for every single droplet, each time from $t=0$ to the present time, involves not only linearly increasing computation time for each time interval, but also impossibly large storage, since one would need to store, for every particle, its acceleration and also that of the fluid at the location of the particle at all past instants. We proposed a novel approach \cite{Prasath2019} to solving this problem. The main advance provided by this approach is that memory costs are drastically reduced, and the memory burden is constant in time. It also involves significantly lower time costs. The reader is referred to \cite{Prasath2019} for more details, but the basic idea is explained here. For several simple flow situations, our approach affords explicit analytical solutions to the MR equation which were unknown hitherto, and a numerical method for the general case. This was made possible by noticing that the term within the square brackets in the MR equation \eqref{mr}, is just the negative of the half (fractional) derivative in time of $\mq \equiv \mv-\mu$. Following the idea of Vishal Vasan, we now write a heat equation in a fictional coordinate $x$, as
\begin{equation}
\mq_t \equiv   \mq_{xx},
\label{heat}
\end{equation}
so the MR equation \eqref{mr} may be rewritten simply as
\begin{equation}
  \dot \mv  = -\frac{\mq}{St} + \frac{3}{(2 \beta St)^{1/2}}\ \mq_x + \frac{1}{Fr^2} \mathbf{e}_g.
    \label{mr_heat}
\end{equation}
Note that equation \eqref{mr_heat} is valid only on the $x=0$ boundary of equation \eqref{heat}, and forms its boundary condition.
An excellent case for study here is that of
a particle sedimenting under gravity in a quiescent ambient. Equation \eqref{exponential_vt} showed how a particle attains its terminal velocity, given by \eqref{terminal}, within a few particle time-scales. In \cite{Prasath2019} on the other hand, we show how the BBH force alters the particle's behaviour in a fundamental way. By our approach, we can write down the complete analytical solution of \eqref{mr_heat}, which includes the \tb{effects of the BBH}, for this problem. For long times, the solution may be expressed as
\begin{equation}
  \mv  = \mvt - 3 \mv_t \sqrt{\frac{St}{2 \beta \pi t}} + O(t^{-3/2}).
    \label{bbh_vt}
\end{equation}
\tb{Accounting for the BBH term} does not change the terminal velocity, but the particle approaches it  algebraically, i.e., slowly, rather than by the exponentially vanishing transients we obtained without history.

\begin{figure}
    \centering
    \includegraphics[width=0.46\columnwidth]{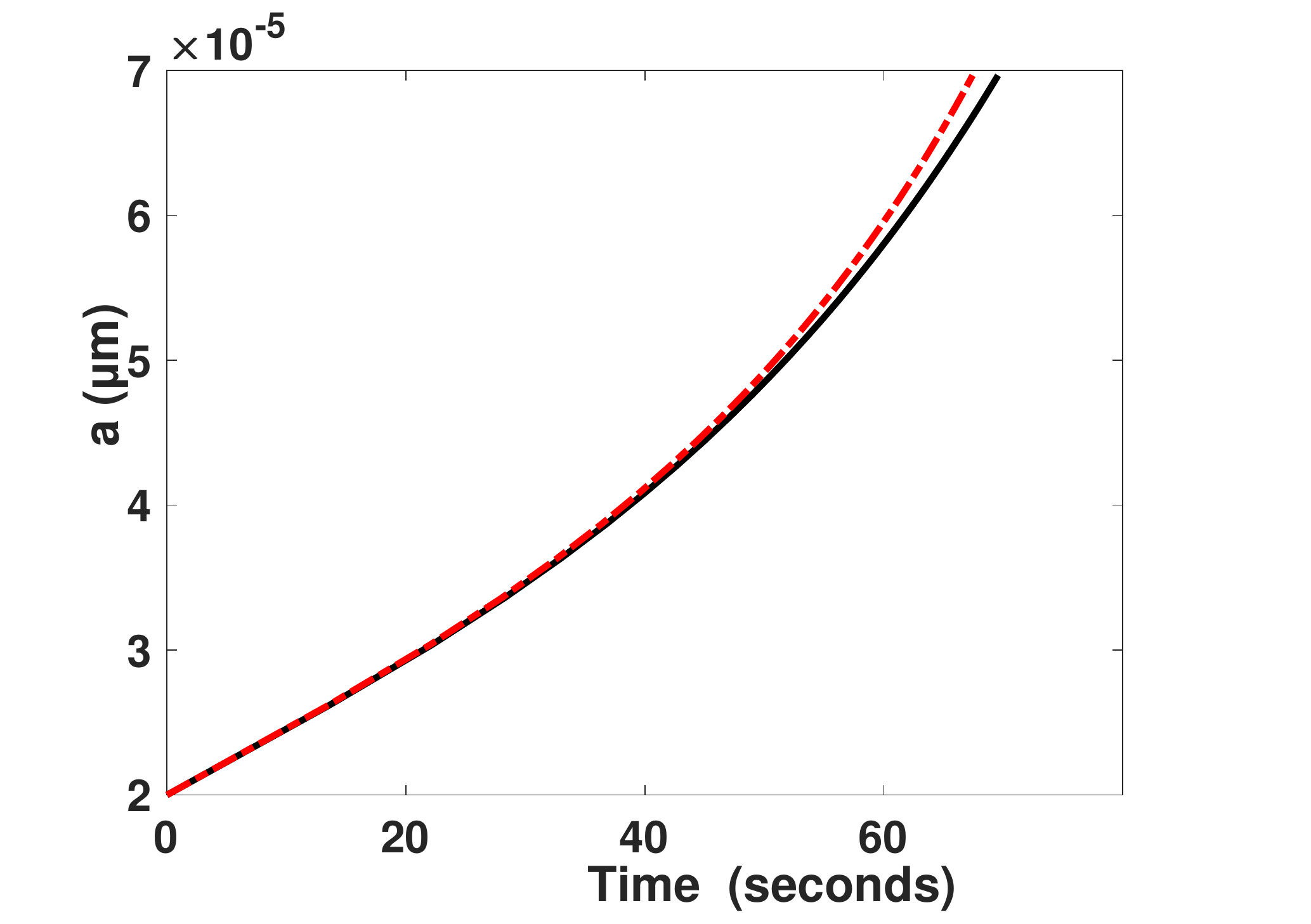}
    \caption{\label{basset} \tb{Growth in time of an initially larger droplet as it sediments through a uniform distribution of smaller droplets, accreting them as it does so. Black solid line: with Basset history. Red dashed line: without Basset history. }}
\end{figure}

\tb{In particular, in \cite{Prasath2019} we solved the complete MR equation \eqref{mr_heat} to obtain the effect of BBH on the growth by collision and coalescence of a larger droplet sedimenting through a sea of identical smaller droplets. The particle density ratio $\beta$ chosen there, $O(10)$, was relevant to other settings of droplet-laden flows, but not directly to clouds, where $\beta=10^3$. For a dilute suspension, the inter-collision time is large enough that we may assume equation (\ref{bbh_vt}) to be valid. The result for $\beta=10^3$ is plotted in figure \ref{basset} and compared against the estimate given by equation \eqref{agrow}.}
\tb{Thus, while neglecting the Basset history term may in general lead to incorrect results, Fig. \ref{basset} shows that for the small droplet number densities and large density ratios typical of clouds, the BBH term is indeed negligible, and the algebraic growth of settling velocity (Eq. \ref{bbh_vt}) does not affect the result.}

\section{Inertia--Thermodynamics coupling}
\label{sec:cloud}

Thus far, we have concerned ourselves with the influence of the flow on the droplets, and neglected the influence of the droplets on the flow. Typically, neglecting the momentum coupling due to the droplets on the flow is justifiable when the volume fractions are sufficiently small. In clouds however, although the volume fraction is only $O(10^{-6})$, droplets can have a profound effect on the flow. This is because droplets in clouds also act as nuclei for condensation, and the accompanying heat release and the consequent buoyancy modification cannot be neglected.

\begin{figure}
    \centering
    \includegraphics[width=0.45\columnwidth]{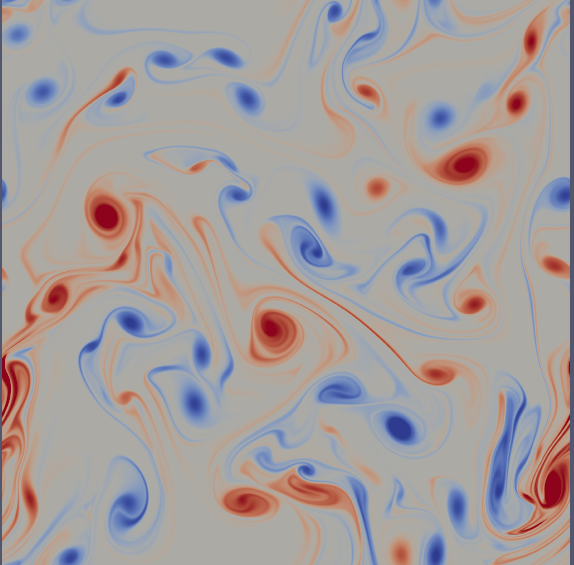}
    \includegraphics[width=0.45\columnwidth]{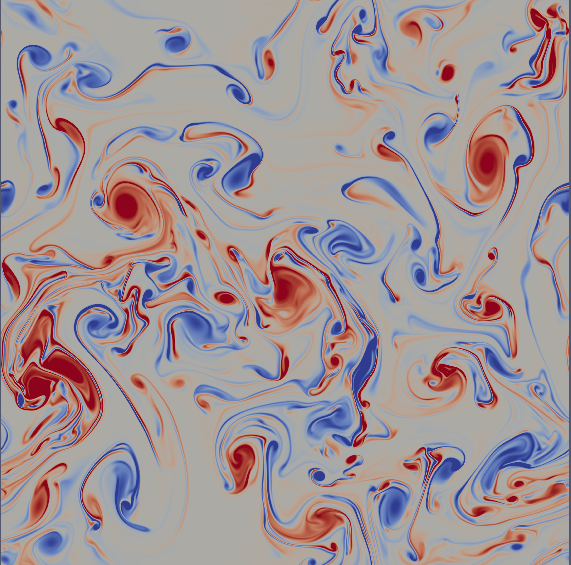}
    \caption{\label{fig:gen_small_scales} The vorticity at $t=50$, \tb{similar to figures 16 and 17 of \cite{Ravichandran2017b}. The domains are $20\times20$ in the $x$ and $y$ directions, and vorticity contours are plotted between the values $(-1,1)$. The nondimensional Reynolds number in the simulations is $Re=2\times10^4$, the phase-change timescale is $St_s = |\omega| \tau_s = 200$ and the droplet Stokes number is $St = |\omega| \tau=1$, where $|\omega|$ is the vorticity scale of the flow, and $\tau_s$ and $\tau$ are the timescales of condensation and droplet inertia respectively. The simulations were performed using a pseudospectral method in the $\omega-\psi$ formulation in two dimensions, with $2/3$rds de-aliasing and a second order exponential time-differencing (ETD-2) scheme. The details may be found in \cite{Ravichandran2017b}.}}
\end{figure}

Droplets, as we have seen, are evacuated from the vicinity of strong vortices. Since droplets also act as condensation nuclei, due to their evacuation, we expect practically no condensation to occur in the void regions surrounding strong vortices in a supersaturated parcel (such as a rising cumulus cloud). What happens in a supersaturated environment when most CCN evacuate the vicinity of strong vortices is that condensation takes place selectively outside these voided spaces, and latent heat is released outside, and the regions occupied by droplets becomes warmer. So strong vortices are now situated within cold patches which are about the size of the caustics radius. If the temperature differences within and outside the patches are large enough, the flow changes qualitatively due to the presence of droplets. This is because buoyancy effects kick in, and baroclinic torque is created. As a result, such flows with phase change display a larger amount of kinetic energy in the small scales than turbulent flows without. The increase in kinetic energy is due to the conversion by baroclinic torque of the increased potential energy due to temperature inhomogeneity, which in turn is due to selective condensation outside void regions. A sample result is shown in figure \ref{fig:gen_small_scales}, which is taken from \cite{Ravichandran2017b}. There, we showed that this effect can occur if the vortex strengths and droplet sizes are such that the product of the inertial Stokes number $St_v$ and the condensation Stokes number $St_s$ is sufficiently large. In a model simulation consisting of just two vortices, we found that if $St_vSt_s >\sim 10^2$, the flow with phase change is qualitatively different from the flow without. The reader is referred to \cite{Ravichandran2017b} for a scaling argument for why this product is important. Intuitively, we may argue that if the condensation time scale is too small, condensation will be complete before droplets have a chance to centrifuge out of vortices. On the other hand, if the droplet time scale is too small, their inertia will be too weak to create void spaces. We may now make an estimate, using data from very large simulations, whether this mechanism can operate in typical conditions in cumulus clouds in the Earth's atmosphere.

The typical inertial timescale in cumulus clouds, where the typical droplet has a radius of $O(10)$ micron, is $O(10^{-3})$s. The timescale for phase change depends on the concentration of droplets and weakly on the ambient temperature, and can be estimated to be $O(10)$s, using
$$
\tau_s = \frac{C \rho_s^0}{4 \pi N a},
$$
where $C \approx 10^7 m\cdot s kg^{-1}$ is a thermodynamic constant, $N$ is the number density of droplets of radius $a$, and $\rho_s^0$ is the saturation vapour density at some base temperature $T_0$. Thus, the product of these timescales is $O(10^{-2})$s$^{2}$. The Kolmogorov timescale, for a dissipation rate of $\epsilon=10^{-2} m^2 s^{-3}$, typical of cumulus and cumulonimbus clouds \cite{Grabowski2013}, is $\tau_k = O(10^{-3/2})$s, giving the Kolmogorov `vorticity scale' $\omega_k = 10^{3/2} s^{-1}$. The product of the inertial and phase-change Stokes numbers is 
$$
St_p \times St_s = \omega_k^2 \tau_p \tau_s = O(10).
$$
Thus, the mechanism in \citep[][]{Ravichandran2017b}, of flow dynamics being changed due to droplets, is not expected to operate for typical Kolmogorov vortices. However, intermittent vortices, we have seen, have vorticities orders of magnitude larger, and thus the product above will be sufficiently large. And since the void fraction becomes significant only at high Reynolds numbers, this is an additional requirement. Our study points to the creation of a larger fraction of smaller and stronger vortices (increased intermittency) if this mechanism were to be in operation.
Since, as we have argued, the primary requirement for the mechanism to operate is the strong vortices associated with intermittent behaviour, this modification of the turbulence is self-sustaining. Explicit evidence for this is the subject of an ongoing study.

Before we end, we mention how phase change can bring about two beautiful cloud formations: mammatus and asperitas, described in \cite{Ravichandran2020mammatus} and \cite{Ravichandran2022asperitas} respectively. As small droplets sediment below the clouds, evaporation can create cooling and a denser layer of air can form just below the clouds. Once this layer becomes pronounced enough, Rayleigh-Taylor instability can take place, giving rise to startling lobe formations, which are called mammatus clouds. When shear is added to this mix, we get a more disordered cloud formation with sharper structures below the cloud base, with some alignment in the shear direction. These are the rare and recently notified asperitas clouds. 

\section{Summary, conclusions, and discussion about future work}

In this article we have examined the consequences of inertial particles (or droplets) centrifuging out of vortical structures in turbulent flow. We have restricted our attention to small spherical droplets or particles in the unsteady Stokes limit in dilute suspension, whose dynamics may be described by the Maxey-Riley equation where Faxen correction terms have been neglected. Given the increasing evidence that turbulence at the small scale contains a significant fraction of tubelike vortices we have treated the Lamb-Oseen vortex as the building block of turbulence, and written down, in \eqref{uni_r} and \eqref{uni_zeta}, the MR equations for a particle in this flow. When taken to the point vortex limit, these equations can be shown to support a boundary-layer structure. Single-vortex caustics can form only when $r_d<r_{cd} \simeq (\Gamma \tau)^{1/2}$. Thus, caustics formation is only relevant for vortices whose size $r_v < r_c$, which we have shown here to consist of vortices smaller than the Kolmogorov scale. \tb{Our arguments do not change significantly by vortex stretching, as in the case of Burgers vortices where we have shown that greater stretching rates lead to faster caustic formation \cite{Agasthya2019}. This is important, given the prevalence of vortex stretching in turbulent flows \citep[][]{Ohkitani2002,Hamlington2008}.}

The connection we find between caustics formation and vortex evacuation enables us to calculate void fractions in the flow, made up of circular (cylindrical in three dimensions) patches around strong vortices which are devoid of particles. Evacuating particles have a much bigger chance of colliding with other particles at short times and, in the case of droplets, coalesce and grow. Using turbulence data at very high Reynolds numbers in conjunction with these ideas, we make the audacious propositions that single-vortex caustics could be important in rain initiation, and that rain initiation needs $R_\lambda>10^3$.

\tb{We then showed, extending the computation of \cite{Prasath2019} to a density ratio $\beta=10^3$ representative of clouds, that the omission of the \tb{Basset-history term}  is justifiable. While neglecting the BBH term alters the manner in which terminal velocity is attained by a falling droplet is algebraic rather than exponential, the effect of this for dilute suspensions with cloud-like density ratios is small.} Finally, we describe how cloud turbulence, due to phase change, can be different from turbulence in other situations.

Several leads for future work emerge from this discussion. First of all, larger and larger Reynolds number direct numerical simulations, which solve the Navier-Stokes equations along with droplet dynamics and the thermodynamics of phase change, are urgently needed, to understand rain initiation in clouds and many other basic physical processes in clouds. Since it is unimaginable that such simulations can include the dynamics of each droplet, a clever superdrop method which includes \tb{the physics of caustics} is crucial to develop. 
\tb{We note that there are many other flow situations where BBH will be important, and our conclusion above is only applicable to particles which are far denser than the fluid, and moreover in dilute suspension. Indeed, the effects of the history term are striking when the density ratio is smaller.} For example, carbonaceous matter sinking through the ocean has density comparable to the background fluid, and correct estimates of carbon sequestration by this process, and its part in mitigating climate change, could be important.

The formation of different cloud shapes contains, apart from breathtaking beauty, much physics, including hitherto undescribed instabilities, and again this area is just opening up for studies. Our findings are also relevant in sprays and industrial flows, and the effects of zones devoid of particles in such flows need to be studied. 

\begin{acknowledgments}
Many thanks are due to Alain Pumir for sharing some data and clarifying some aspects of their recent study, which forms a crucial input for some of our arguments.  RG gratefully acknowledges support of the Department of Atomic Energy, Government of India, under project no. RTI4001. SR was supported at Nordita under the Swedish Research Council grant no. 638-2013-9243. Nordita is partially funded by Nordforsk.
\end{acknowledgments}

%

\end{document}